\documentclass[prd,showpacs,superscriptaddress,twocolumn,floatfix,preprintnumbers,altaffilletter, letterpaper]{revtex4-1}

\usepackage{amsmath,amssymb}  % for math

\usepackage{xcolor}
\usepackage{graphicx} % needed for figures
\usepackage{subcaption}
\usepackage{multirow}
\usepackage{enumitem}

\usepackage[colorlinks=true,citecolor=teal, linkcolor=blue]{hyperref}

\usepackage[T1]{fontenc}

\begin{document}

\title{Bilinear noise subtraction at the GEO\,600 observatory}

\author{N. Mukund} 
\email{nikhil.mukund@aei.mpg.de}
\affiliation{Max-Planck-Institut f{\"u}r Gravitationsphysik (Albert-Einstein-Institut) and Institut f{\"u}r Gravitationsphysik, Leibniz Universit{\"a}t Hannover, Callinstra{\ss}e 38, 30167 Hannover, Germany} 

\author{J. Lough} 
%\email{james.lough@aei.mpg.de}
\affiliation{Max-Planck-Institut f{\"u}r Gravitationsphysik (Albert-Einstein-Institut) and Institut f{\"u}r Gravitationsphysik, Leibniz Universit{\"a}t Hannover, Callinstra{\ss}e 38, 30167 Hannover, Germany} 

\author{C. Affeldt} 
%\email{christoph.affeldt@aei.mpg.de}
\affiliation{Max-Planck-Institut f{\"u}r Gravitationsphysik (Albert-Einstein-Institut) and Institut f{\"u}r Gravitationsphysik, Leibniz Universit{\"a}t Hannover, Callinstra{\ss}e 38, 30167 Hannover, Germany} 

\author{F. Bergamin} 
\affiliation{Max-Planck-Institut f{\"u}r Gravitationsphysik (Albert-Einstein-Institut) and Institut f{\"u}r Gravitationsphysik, Leibniz Universit{\"a}t Hannover, Callinstra{\ss}e 38, 30167 Hannover, Germany}

\author{A. Bisht} 
\affiliation{Max-Planck-Institut f{\"u}r Gravitationsphysik (Albert-Einstein-Institut) and Institut f{\"u}r Gravitationsphysik, Leibniz Universit{\"a}t Hannover, Callinstra{\ss}e 38, 30167 Hannover, Germany}

\author{M. Brinkmann} 
\affiliation{Max-Planck-Institut f{\"u}r Gravitationsphysik (Albert-Einstein-Institut) and Institut f{\"u}r Gravitationsphysik, Leibniz Universit{\"a}t Hannover, Callinstra{\ss}e 38, 30167 Hannover, Germany}

\author{V. Kringel} 
\affiliation{Max-Planck-Institut f{\"u}r Gravitationsphysik (Albert-Einstein-Institut) and Institut f{\"u}r Gravitationsphysik, Leibniz Universit{\"a}t Hannover, Callinstra{\ss}e 38, 30167 Hannover, Germany}

\author{H. L{\"u}ck} 
\affiliation{Max-Planck-Institut f{\"u}r Gravitationsphysik (Albert-Einstein-Institut) and Institut f{\"u}r Gravitationsphysik, Leibniz Universit{\"a}t Hannover, Callinstra{\ss}e 38, 30167 Hannover, Germany}

\author{S. Nadji} 
\affiliation{Max-Planck-Institut f{\"u}r Gravitationsphysik (Albert-Einstein-Institut) and Institut f{\"u}r Gravitationsphysik, Leibniz Universit{\"a}t Hannover, Callinstra{\ss}e 38, 30167 Hannover, Germany}

\author{M. Weinert} 
\affiliation{Max-Planck-Institut f{\"u}r Gravitationsphysik (Albert-Einstein-Institut) and Institut f{\"u}r Gravitationsphysik, Leibniz Universit{\"a}t Hannover, Callinstra{\ss}e 38, 30167 Hannover, Germany}

\author{K. Danzmann} 
\affiliation{Max-Planck-Institut f{\"u}r Gravitationsphysik (Albert-Einstein-Institut) and Institut f{\"u}r Gravitationsphysik, Leibniz Universit{\"a}t Hannover, Callinstra{\ss}e 38, 30167 Hannover, Germany}

\date{\today}

\begin{abstract}

Longitudinal control signals used to keep gravitational wave detectors at a stable operating point are often affected by modulations from test mass misalignments leading to an elevated noise floor ranging from 50 to 500 Hz.  Nonstationary noise of this kind results in modulation sidebands and increases the number of glitches observed in the calibrated strain data.  These artifacts ultimately affect the data quality and decrease the efficiency of the data analysis pipelines looking for astrophysical signals from continuous waves as well as the transient events. In this work, we develop a scheme to subtract one such bilinear noise from the gravitational wave strain data and demonstrate it at the GEO\,600 observatory. We estimate the coupling by making use of narrow-band signal injections that are already in place for noise projection purposes and construct a coherent bilinear signal by a two-stage system identification process. We improve upon the existing filter design techniques by employing a Bayesian adaptive directed search strategy that optimizes across the several key parameters that affect the accuracy of the estimated model. The scheme takes into account the possible nonstationarities in the coupling by periodically updating the involved filter coefficients. The resulting postoffline subtraction leads to a suppression of modulation sidebands around the calibration lines along with a broadband reduction of the midfrequency noise floor.  The observed increase in the astrophysical range and a reduction in the occurrence of nonastrophysical transients suggest that the above method is a viable data cleaning technique for current and future generation gravitational wave observatories.

\end{abstract}

\maketitle

\section{Introduction}

GEO\,600 is a British-German gravitational wave (GW) detector \citep{Willke_2002,LuEA2006} located in Ruthe, Hannover that searches for signals in the audio-band frequencies generated from astrophysical sources such as black holes and neutron stars. It is a dual recycled Michelson laser interferometer \citep{LuEA2010,GEO2016} with 600m long folded arms reaching an average peak sensitivity of $2 \times 10^{-22} \; \text{Hz}^{-1/2}$ at 1 kHz and works in tandem with the global network of detectors that includes Advanced LIGO \citep{Aasi_2015} and Advanced Virgo \citep{Acernese_2014}. GEO\,600 also functions as a GW technology demonstrator and has pioneered the use of several vital technologies \citep{affeldt2014advanced}, which subsequently were incorporated in the larger detectors. These include the use of a squeezed light source \citep{abadie2011gravitational}, monolithic suspension \citep{Goer:2004ppa}, signal recycling \citep{Heinzel_2002}, active thermal compensation \citep{Luck:2004us,Wittel:18}, and electrostatic drive-based actuation. The observatory has demonstrated consistent levels of squeezing over year-long time scales \citep{grote2013first} and can now reduce the shot noise in the kiloHertz regime by a factor close to 2 \citep{lough2020demonstration}. The automated alignment \citep{Grote_2002} and locking \citep{grote2003making} scheme has led to the stable operation of the detector, and over the last several years, the duty cycle has consistently been above $80\%$. Since the sensitivity around 3 kHz is only an order of magnitude worse than Advanced LIGO, the data from the detector was utilized in constraining the properties of the probable postmerger signal from the low mass compact binary inspiral signal GW170817 \citep{PhysRevX.9.011001}. More recent work \citep{PhysRevResearch.1.033187} has also pointed out the relatively higher sensitivity of GEO\,600 compared to other kilometer-scale GW interferometers in searches looking for dark matter field oscillations in the range 100 Hz to 10 kHz. The sensitivity enhancement towards these elusive events is due to the absence of Fabry-Perot arm cavities and the resulting higher bandwidth.

Similar to other GW detectors, noise sources arising from thermal, seismic or quantum mechanical properties of light pose a fundamental limit to the achievable sensitivity. Equally, troublesome are the technical ones arising from the various auxiliary control loops that are used to keep the detector at a stable operating point. The standard strategy adopted, principally for seismic and technical noise sources, is to estimate the linear part of the coupling and subtract it off via online feedforward cancellation \citep{Giaime:2003zf,smith2005feedforward,doi:10.1063/1.3675891,Meadors:2013lja}. Since the strain and auxiliary channel data are stored, it is also possible to perform offline subtraction at a later time. Such offline cleaning has the advantage of preserving the original data, and the appropriate filters can be reestimated for every data segment, thus minimizing the amount of additional noise injection. Solutions based on Wiener filtering are effective in regressing environmental disturbances \citep{Tiwari_2015} and have been applied to tackle correlated magnetic noises arising from Schumman resonances \citep{Coughlin_2018_Schumann} as well as gravity gradients caused by seismic surface waves \citep{10.1007/978-88-470-2113-6_44,Beker_2012,Driggers_2012,Coughlin_2016_NN}. Effectiveness of offline analysis was recently demonstrated in Advanced LIGO second observation run (O2) data, where the removal of laser jitter coupling led to an improvement in the detection range by a factor of $15\%$ \citep{PhysRevD.99.042001}.

The presence of higher-order couplings has previously been shown to impact searches looking for short duration transients. By looking at statistically significant temporal coincidences between strain and bilinear channel combinations, it is possible to veto time segments leading to a reduction in the number of such false triggers \citep{PhysRevD.89.122001}. When no prior information is available regarding the nonlinear nature in which various signals are combined, higher-order statistics-based quantities like quadratic phase coupling have been suggested \citep{bose:2016sqv} as a useful metric to identify the involved pair. A commonly observed bilinear pair consists of a slowly changing alignment degree of freedom and a rapidly varying one used for controlling the length of an optical cavity. In this work, we look into a form of bilinear coupling arising from the longitudinal control of the signal recycling mirror and show how accurate system identification can help in time domain subtraction of these from the calibrated strain data. The primary motivation towards this work arose mostly from the observation of significant sidebands around some of the narrow-band signal injections that are used to estimate the noise contributions arising from various degrees of freedom. Our motivation also stems from the observation of the midfrequency (50-500 Hz) noise at GEO\,600, which goes up with the increase of input laser power. Past attempts to identify the source of this noise ruled out linear couplings and consequently pointed towards effects from higher-order couplings.
 
 The rest of the paper is organized as follows: Sec. \ref{sec:methods} presents our understanding of the noise coupling mechanism while Sec. \ref{sec:sysID} talks about challenges encountered in the accurate system identification and the methods adopted to overcome them. Postsubtraction results are provided in \ref{sec:results}, where we also talk about the effect on certain data quality metrics such as glitch rate, and improvements to the astrophysical range. Finally we present our conclusions in Sec. \ref{sec:conclusions}.

\section{Coupling Mechanism}\label{sec:methods}

\iffalse
\begin{figure}[!htb]
  \centering
    \includegraphics[width=\linewidth]{FIG1}
  \caption{Longitudinal control scheme of GEO\,600 \citep{grote2003making}}
  \label{fig:lengthControlSchematic}
\end{figure}
\fi
Dual recycling at GEO\,600 consists of power recycling (PR) to increase the circulating carrier field and signal recycling (SR) for resonant enhancement of the signal sidebands. Depending on the respective finesse of PR and SR cavities, this dual recycling enables the detector to have different storage time for carrier and storage signals. We can describe the frequency-dependent sensitivity of an interferometer in terms of the transfer function between the incident gravitational wave signal and the output photocurrent at the photodetector. For a dual recycled Michelson configuration, this response (normalized in terms of input power) is given as
\citep{mizuno1995comparison},
\begin{equation}
G(\omega) = \sqrt{\mathcal{G}_{arm}} \times \frac{-\tau_{s}}{1 - \rho_{s} \rho_{a} e^{-i \omega t_{r}}} \times \frac{\rho_{a} \omega_{0}}{4} \frac{1 - e^{-i \omega t_r}}{i \omega}  \,
\end{equation}
where $\mathcal{G}_{arm}$ is the ratio of power injected to that circulating in both the arms; $\tau_{s},\rho_{s} \& \rho_{a}$ are the respective amplitude transmittance and reflectivities of the mirrors that form the signal recycling cavity; $\omega_{0}$ is the laser frequency and $t_{r}$ is the retarded time that incorporates the multiple reflections within the arm. To keep the detector at a stable operating point and to maintain resonance in cavities, it is necessary to keep the longitudinal motion of the key optics limited to a  fraction of the carrier field wavelength. We achieve this using multiple servo loops, of which three are used for controlling the Michelson differential arm length as well as the length of PR and SR cavities \citep{grote2003making}.  The interferometer calibration process \citep{hewitson2003calibration,Hewitson_2003,Leong_2012} refers to the time domain reconstruction of the test mass displacement based on the data obtained from multiple channels of varying frequency responses and signal-to-noise ratios (SNRs). For frequencies above the unity gain of the Michelson servo, we obtain the optical gain by observing the detector output for a known amount of perturbation applied at the end mirrors.  These injections of known length modulations applied at specific frequencies are commonly referred to as calibration lines. Often such lines are added to various control signals to get an estimate of their respective contributions to the GW sensitive signal. Specifically, for the case of SR cavity length (SRCL) control, this addition happens digitally to the feedback signals before they are sent to the coil-magnet actuators. The SR mirror, in particular, is suspended via two-stage pendulum to damp the horizontal motion and two-stage vertical cantilever springs to suppress the vertical disturbances. To create a longitudinal control signal for the SR mirror, we pick off light from the antireflection coated side of the beamsplitter and demodulate it at the Schnupp modulation sideband frequency of 9.18 MHz. Actuation is then carried out through a separate reaction chain using three magnet-coil actuators situated at its lowest stage.  
\begin{figure}[!htb]
  \centering
    \includegraphics[width=\linewidth]{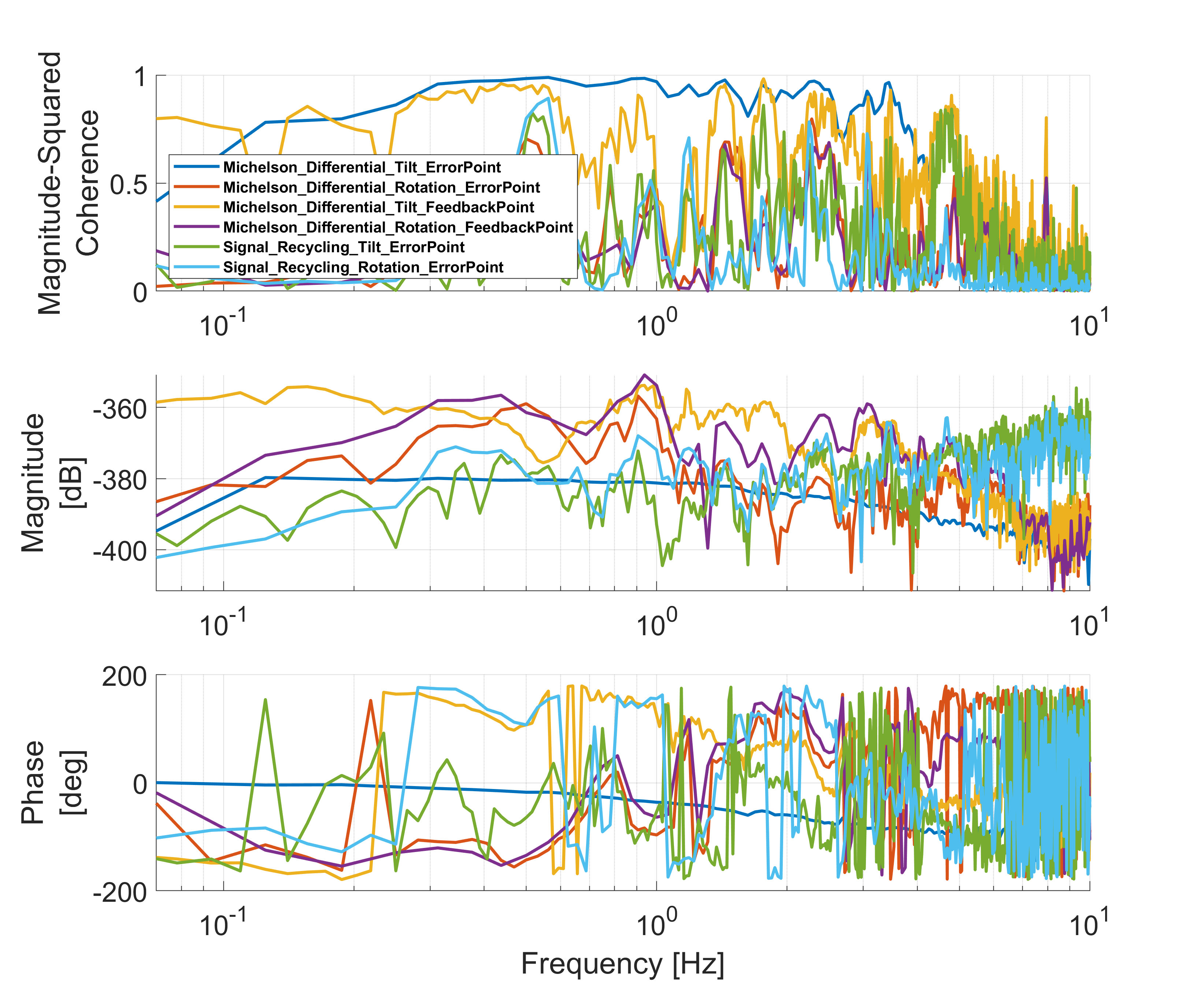}
  \caption{Linear coupling between prominent alignment degrees of freedom and GW strain signal demodulated at 320 Hz. }
  \label{fig:coherencePlots}
\end{figure}
\begin{figure}[!htb]
\centering
    \includegraphics[width=0.9\linewidth]{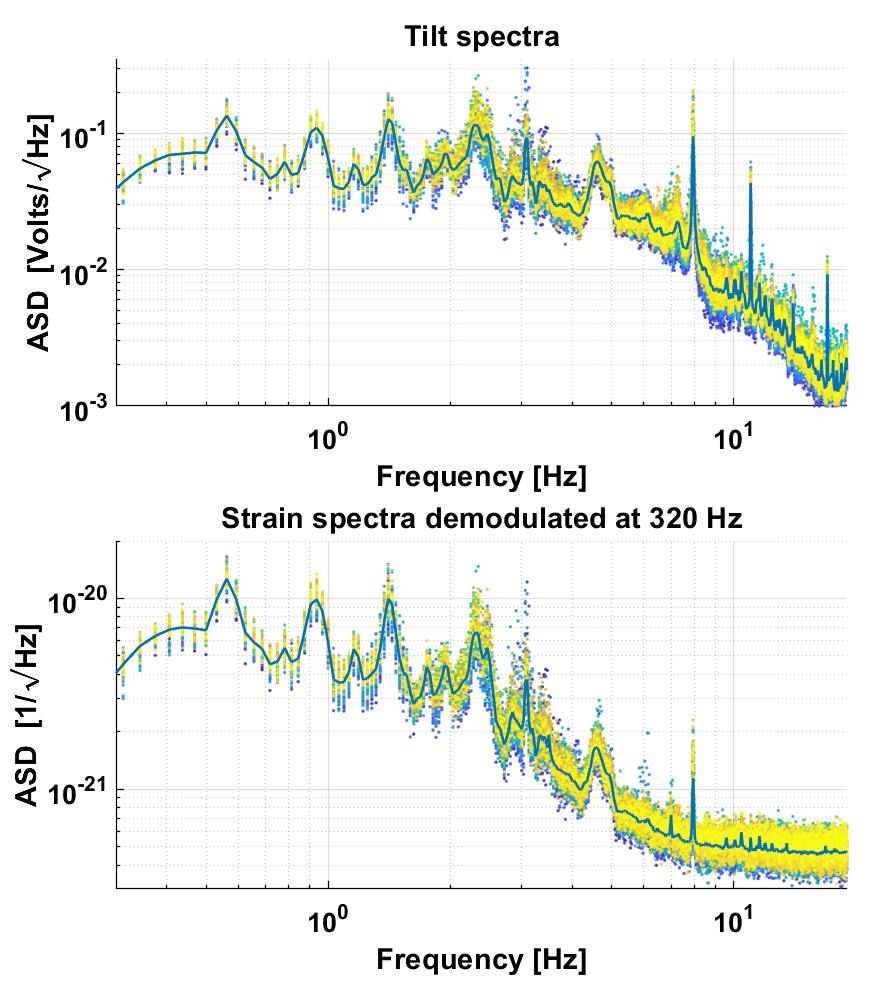}
  \caption{ Top plot shows spectra for differential tilt measured between end test mass mirrors, and the bottom one shows the GW strain signal demodulated at the signal-recycling calibration line at 320 Hz. Each of the dotted lines is constructed using 8 minute long segments, and the solid line gives the median average for 12 hours of data.}
  \label{fig:spectral_variation}
\end{figure}
 \begin{figure*}[!htb]
  \centering
  \includegraphics[width=0.95\linewidth]{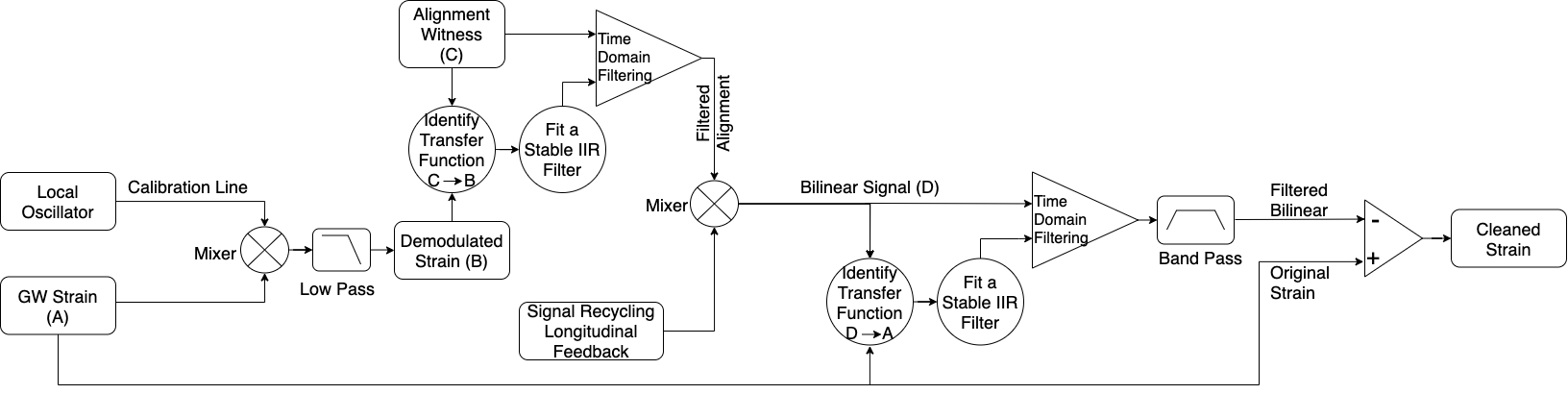}
  \caption{Flow graph depicting the steps involved in the construction of a coherent bilinear signal from auxiliary channels.}
  \label{fig:bilinearFlowGraph}
\end{figure*}
The SRCL control loop has a bandwidth of 35 Hz and is the strongest known technical noise source up to 200 Hz \citep{Wittel_SRCL}. The loop shape is sensitive to factors such as the alignment, circulating power, and other unknown parameters that change with time. The SRCL noise coupling to Michelson differential arises majorly from the small offset introduced between the end mirrors ($2.10 \times 10^{-10}$ m), which leaks out a tiny amount of carrier light to the dark port to function as a local oscillator for the DC readout scheme \citep{Izumi_2016,vaishali_thesis}. %Efforts to suppress the out-of-bandwidth noise by making the controller transfer function roll-off steeper via low pass filtering adds additional complexity as well as increased instability.

Residual angular motion on any of the suspended optics can induce phase modulations in the circulating carrier field, leading to higher-order coupling in the strain data. Often, one of the first signs for such a coupling is the presence of sidebands seen around the narrow-band line injections. In particular, for GEO\,600, the slight offset between the point of suspension and the horizontal axis of the test mass mirrors results in an enhanced length to angle coupling along with the tilt as compared to the rotational degree of freedom. We correct for this differential tilt by applying appropriate feedback at the end mass mirrors, and by locally damping the folding mirrors. Additionally, as the corrective forces are applied at the upper stage of the suspended test mass, the corresponding feedback transfer function gets added complexities making it harder to model compared to the one obtained from the error point measurements. Figure \ref{fig:coherencePlots} compares the linear coupling measured from prominent alignment degrees of freedom to the GW strain signal demodulated at the signal-recycling calibration line frequency of 320 Hz. As the Michelson tilt error point signal provides maximum coherence with demodulated strain, we select it for the further analysis described in this paper. The multiple peaks that we see in the demodulated strain match well with resonant modes of the test mass suspension tilt spectra (see Fig.  \ref{fig:spectral_variation}), further strengthening the possibility of the coupling mentioned above. In reality, angular motion is imprinted in the strain signal in a broadband sense. However, we prominently witness it around these calibration lines because of their higher SNR. These sidebands often vary in strength on time scales of a few tens of minutes and are known to increase the overall transient noise level. 
 \\

These non-Gaussian transients, as well as other environmental disturbances, elevate the nonstationary level leading to false triggers in template-based and unmodeled burst search pipelines. Several kinds of veto techniques have been developed within the GW data analysis groups to identify these glitches and minimize their impact on the significance of real events \citep{slutsky2010methods}. Such vetoes could be based on statistically significant temporal coincidences between the strain and auxiliary data channels \citep{Smith_2011} or be based on events seen in null-stream constructed out of the two calibrated strain quadratures \citep{Hewitson_2005}. For template-based searches, a $\chi^{2}$ time-frequency discriminator is often used to check for consistency between the trigger and the expected event \citep{PhysRevD.71.062001}. More recently, techniques based on unsupervised and supervised machine learning have also been successful in identifying several glitch class populations observed within the strain data \citep{2013PhRvD..88f2003B,2015CQGra..32u5012P,PhysRevD.95.104059,Zevin:2016qwy}. As compared to transient sources, the sensitivity of GW searches looking for continuous wave signals is more affected by persistent narrow-band spectral lines arising from power line harmonics and calibration lines along with their associated sidebands. These searches require prolonged stretches of data that are often spread across multiple observation periods and so to minimize the contamination, a specific bandwidth (a few Hz) of data around these lines is often removed prior to carrying out the actual analysis. Constructing a coherent bilinear signal based on our understanding of the involved coupling, and carrying out a postoffline subtraction as described in the following section could thus lead to a certain degree of improvement in the various data analysis pipelines.

\section{System Identification } \label{sec:sysID}
Before tackling any nonlinear effects, it is essential to minimize the linear coupling arising from the detuning of the SR mirror. This length offset is usually identifiable from the presence of a more prominent central peak at the injection line as compared to its sidebands. We correct for this offset through feedback actuation in proportion to observed peak amplitude. Figure \ref{fig:bilinearFlowGraph} depicts the procedure adopted for bilinear noise subtraction, and it mainly involves two sequential stages of linear system identification. As described in the previous section, we start with the Michelson differential tilt and model its transfer function to the demodulated strain using a stable IIR filter. The SRCL signal is then multiplied with this filtered tilt to generate an intermediate signal, which is further filtered based on its transfer function to the full strain data yielding the final bilinear signal.
 \begin{figure*}[!htb]
  \centering
    \includegraphics[width=\linewidth]{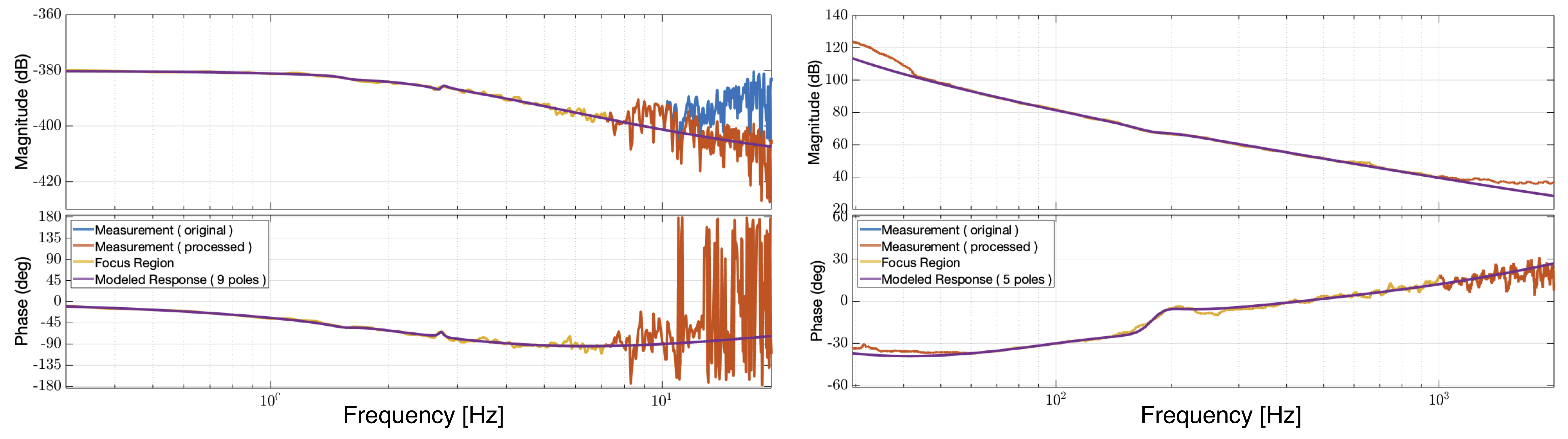}
  \caption{System identification using Bayesian adaptive directed search assisted vector fitting.}
  \label{fig:bodeplots_fits}
\end{figure*}
\begin{figure*}[!htb]
  \centering
    \includegraphics[width=\linewidth]{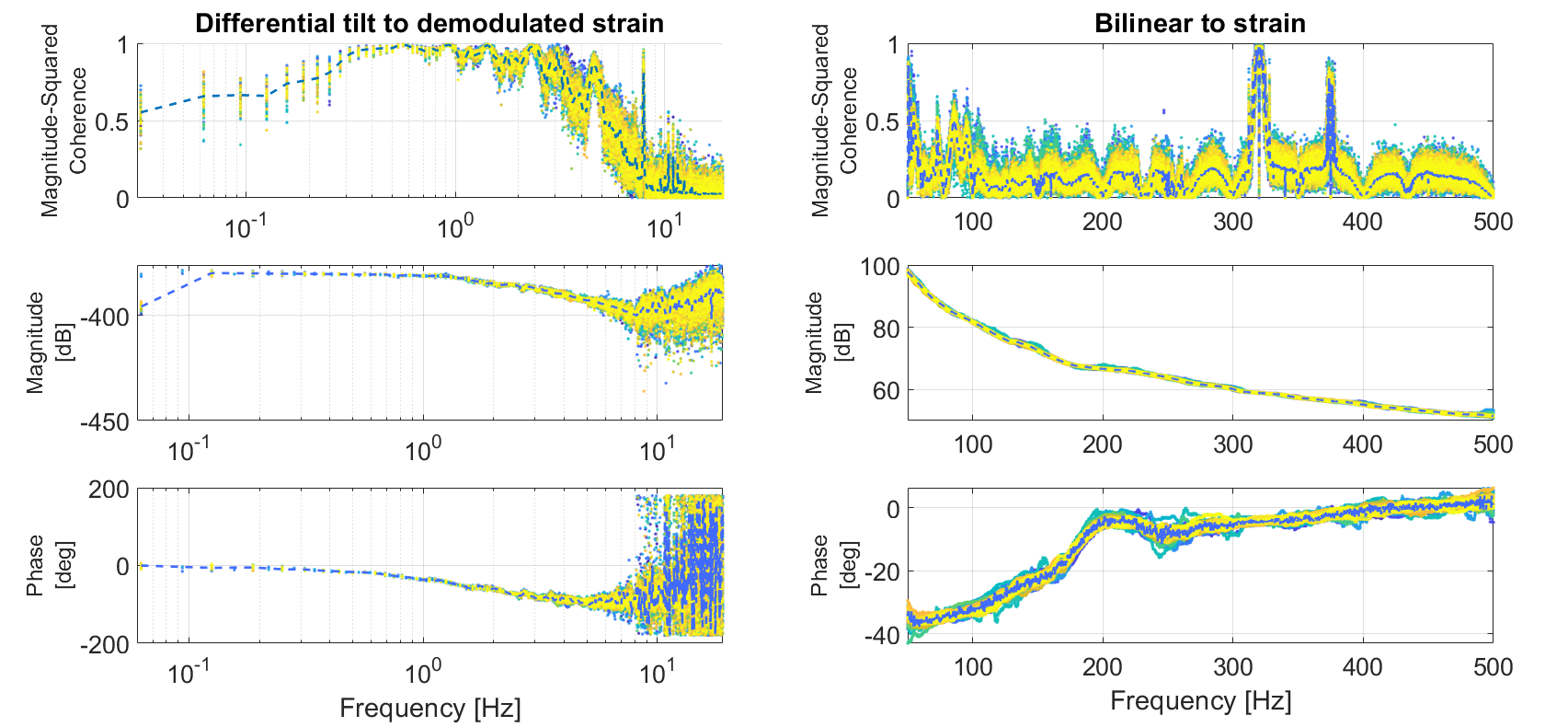}
  \caption{Estimated transfer functions and their temporal variability. The first plot models the coupling between differential tilt and demodulated strain while the second one provides the coupling between the constructed bilinear signal and measured strain signal. The dashed line gives the median average.}
  \label{fig:bodeplots}
\end{figure*}

 We make use of frequency-domain Bode diagrams to identify the linear coupling as these plots quickly reveal the dynamic range and sharp resonances involved within the coupling. As the optimal frequency resolution needed for the fast Fourier transform is not known a priori, we scan across a range of resolutions and choose the one that minimizes the phase jitter across the band of interest. This jitter is quantified using the absolute value of the phase derivative, and a scalar metric useful for comparison is created by integrating it across the midfrequencies. The optimally resolved transfer function is then modeled as an IIR filter, primarily so that the digital control system can keep up and also since it usually provides a better fit with a fewer number of filter coefficients as compared to an equivalent FIR representation. 

 Optimal modeling of the estimated transfer function is critical as any misrepresentations can lead to noise injection. Identifying the right zeros and poles to match the desired response, especially in the presence of noises, can be considered as a form of a nonconvex optimization problem. We express the unknown ZPK model as a rational function,
\begin{equation}  
H(s) \approx \sum\limits_{n=1}^{N} \frac{c_{n}}{s-a_{n}} + d + sh \;,
\label{eq:VF_1}
\end{equation}
where $c_{n}$ and $a_{n}$ can either be real quantities or complex conjugate pairs, while d and h are both real. As the unknown poles $a_{n}$ occur in the denominator, Eq. (\ref{eq:VF_1} ) cannot be directly solved as a linear problem. Using an unknown function $\sigma(s)$ and with an initial set of poles $\tilde{a}_{n}$ and zeros $\tilde{c}_{n}$, the above problem can be transformed into a linear one,
\begin{equation}
\Bigg( \sum\limits_{n=1}^{N} \frac{c_{n}}{s-\tilde{a}_{n}} +d + sh \Bigg) - \Bigg( \sum\limits_{n=1}^{N} \frac{\tilde{c}_{n}}{s-\tilde{a}_{n}} \Bigg) H(s) \approx H(s) \;,
\end{equation}
where
\begin{equation} 
\begin{bmatrix}
\sigma(s) H(s)\\
\sigma(s) \\
\end{bmatrix} 
\approx
\begin{bmatrix}
\sum\limits_{n=1}^{N} \frac{c_{n}}{s-\tilde{a}_{n}} +d + sh\\
\sum\limits_{n=1}^{N} \frac{\tilde{c}_{n}}{s-\tilde{a}_{n}}+1 \;
\end{bmatrix} \;.
\end{equation}
We then make use of a vector fitting (VF) \citep{gustavsen1999rational,gustavsen2006relaxed} scheme to solve the above equation hierarchically. The process is carried out iteratively by first carrying out pole identification followed by the identification of residues, $ c_{n} $. To obtain a better fit without suffering from issues related to ill conditioning, we follow the typical strategy of starting with complex conjugate pole pairs whose imaginary part spans either linearly or logarithmically the frequency span of interest. The routine further enforces stability by forcing all the poles to be on the left half-plane in the continuous Laplace domain. 
\begin{figure*}[!htb]
  \centering
    \includegraphics[width=0.75\linewidth]{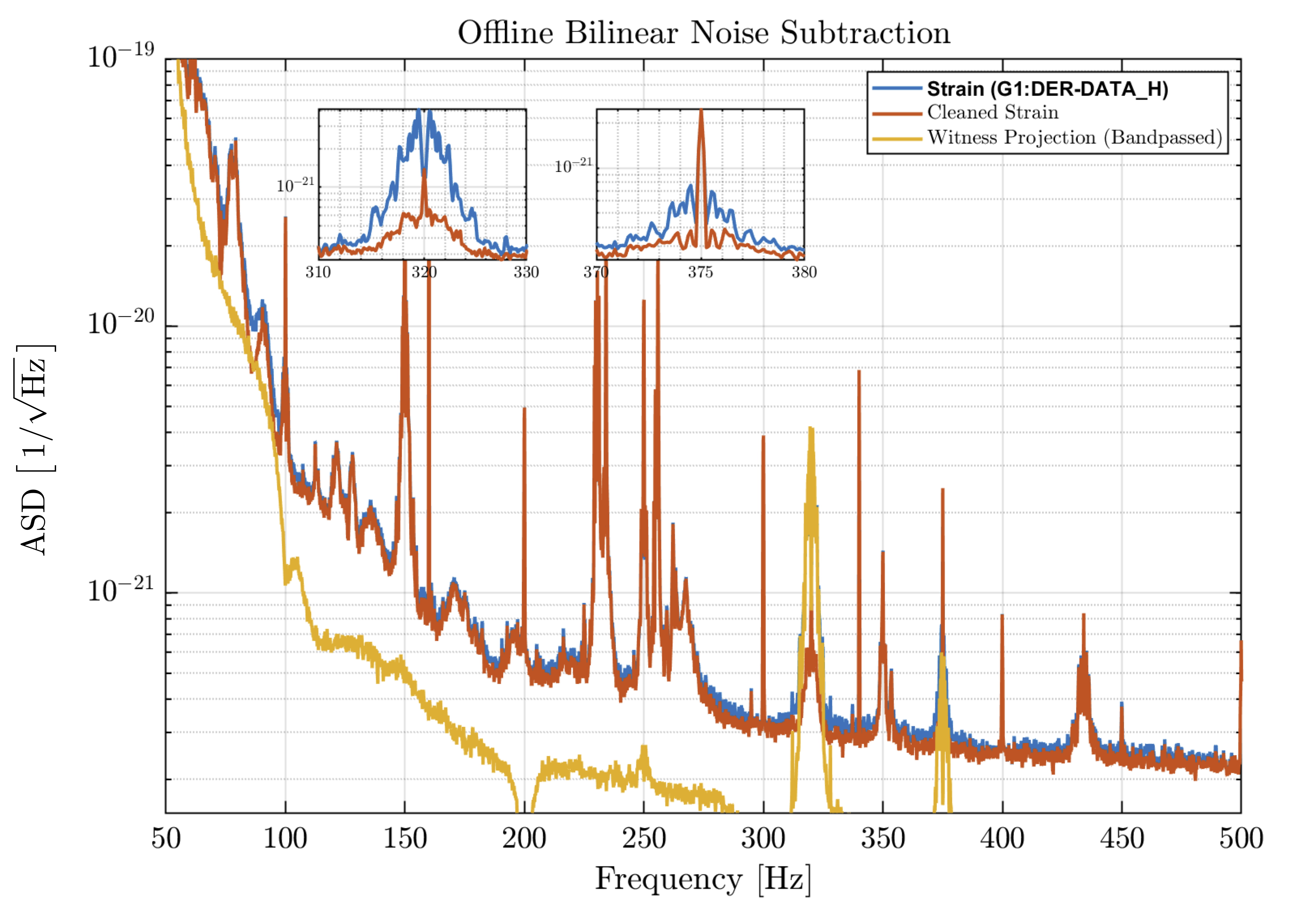}
  \caption{Typical strain spectra that show the effect of noise subtraction from 50 to 500 Hz. Inset plots highlight the sideband suppression obtained around the calibration lines at 320 and 375 Hz. }
  \label{fig:bilinearL_subtraction}
\end{figure*}
\begin{figure*}[!htb]
  \centering
    \includegraphics[width=\linewidth]{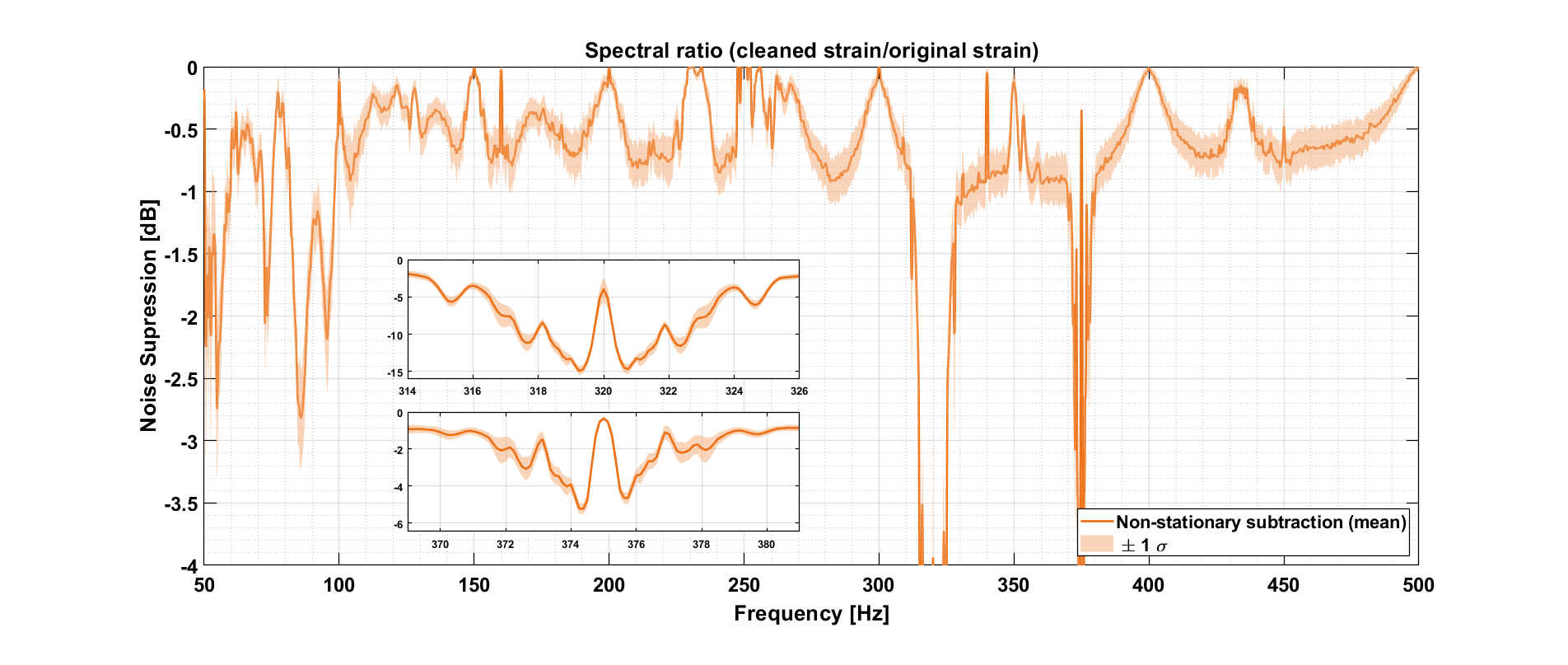}
  \caption{The plot shows the broadband improvements to strain sensitivity expressed in terms of spectral ratio for 12 hours of data starting from 21-09-2015 23:00:00 (UTC). For each segment of 512 s duration, the filter coefficients are updated prior to carrying out the subtraction.}
  \label{fig:bilinearL_subtraction2}
\end{figure*}
The quality of the fit is assessed using the normalized root mean square error metric, 
\begin{equation}
 nrmse(y) = \frac{\|y-\hat{y}\|}{\|y-mean(y)\|} \; ,
\end{equation}
where $y$ and $\hat{y}$ respectively are the measured and modeled response. The initial pole placement, number of poles, frequency-dependent weighting factor, and initial and final frequencies are all unknown factors that have a significant effect on the overall quality of the model. To optimize across these, we make use of Bayesian adaptive directed search (BADS) \citep{acerbi2017practical} to scan across the respective parameter space without much invoking a heavy computational overhead. Bayesian optimization (BO) is typically used in machine learning applications for model fitting, especially when the cost functions are expensive to calculate. Such scenarios are often encountered in hyperparameter tuning but usually come with large algorithmic overhead and require some amount of fine-tuning. The advantage of BADS comes from its use of derivative-free mesh adaptive direct search combined with its use of BO via surrogates based on Gaussian processes, which drastically speeds up the function evaluations. Using the procedure mentioned above, we obtain a reasonably good fit for the estimated transfer functions, as shown in Fig. \ref{fig:bodeplots_fits}. Before performing time-domain filtering, we discretize the continuous domain model at the sampling frequency (16 kHz) and then finally convert it to second-order sections to overcome the issues related to numerical rounding errors \citep{mitra2006digital,jackson2013digital}. \\
 \begin{figure*}[!htb]
  \centering
    \includegraphics[width=0.49\linewidth]{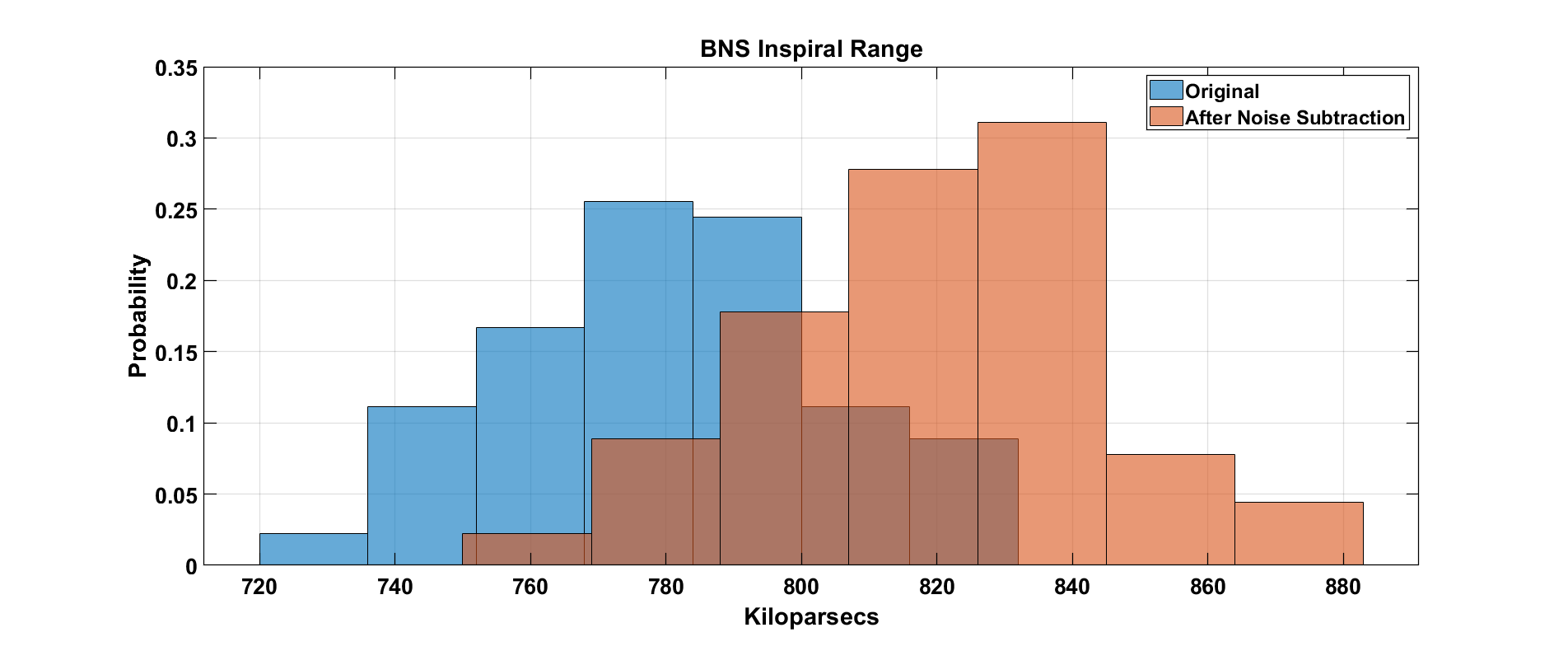}
    \includegraphics[width=0.49\linewidth]{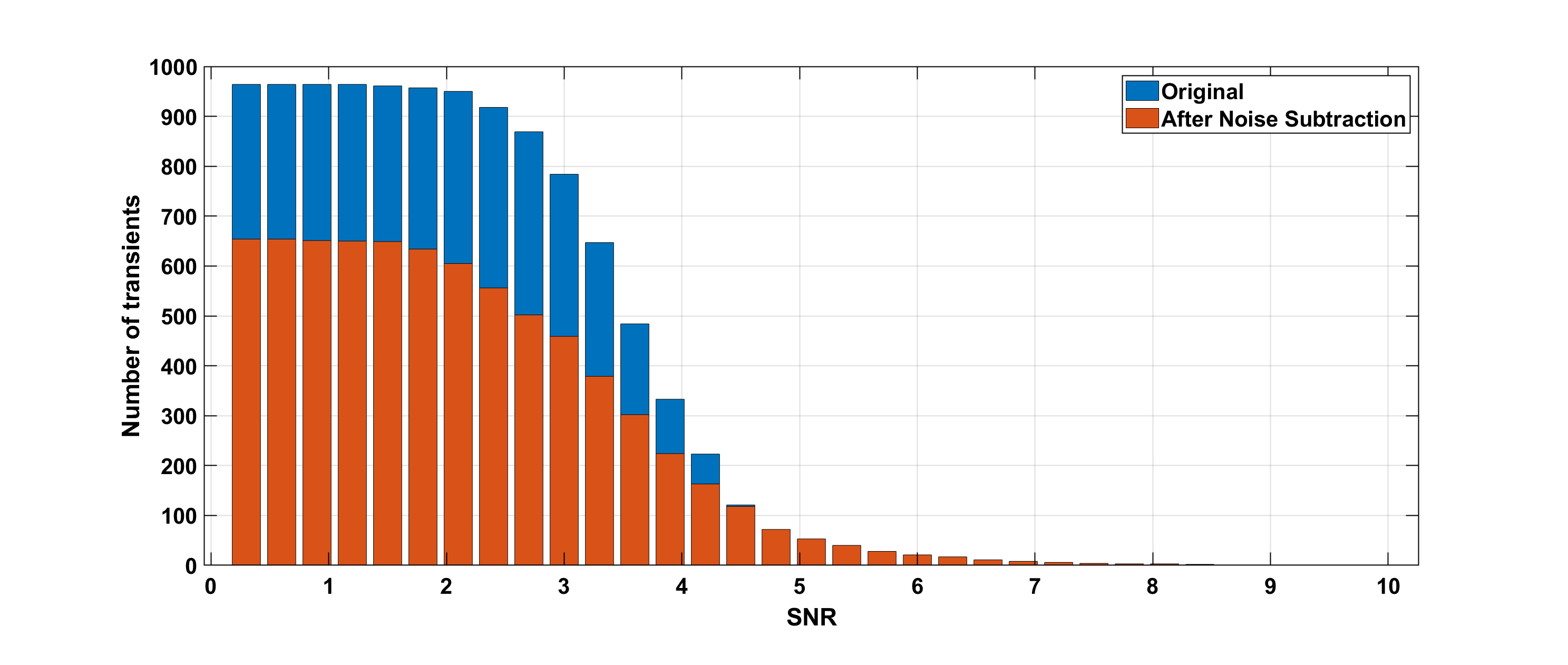}
  \caption{ Improvements in data quality. The first plot shows the increase in horizon distance for an optimally located and oriented binary neutron star of 1.4 solar mass each. The second plot gives the cumulative histogram of excess power transients above a given SNR before and after the bilinear noise subtraction.}
  \label{fig:DQ_improvements}
\end{figure*}
Finally, to achieve optimal subtraction, we address the issue related to the presence of nonstationarities in each of the identified couplings. Updating the filter coefficients to tackle such nonstationarities has recently been shown to provide better subtraction for the case of nonlinear noise observed in the Advanced LIGO detectors \citep{PhysRevD.101.042003}. Often, as a result of the ongoing commissioning activities at the site, the intermediate filters used in the actuation path are very likely to get modified, so re-calculating the filter coefficients is, in general, a desirable strategy. While analyzing the data spread across a period of 12 hours at GEO\,600, we do observe slight variations in the identified transfer functions, as shown in Fig. \ref{fig:bodeplots}. Hence, we carry out periodic system identification every 8 minutes and accordingly carry out the bilinear subtraction using the updated filter coefficients.

\begin{figure*}[!htb]
  \centering
    \includegraphics[width=0.95\linewidth]{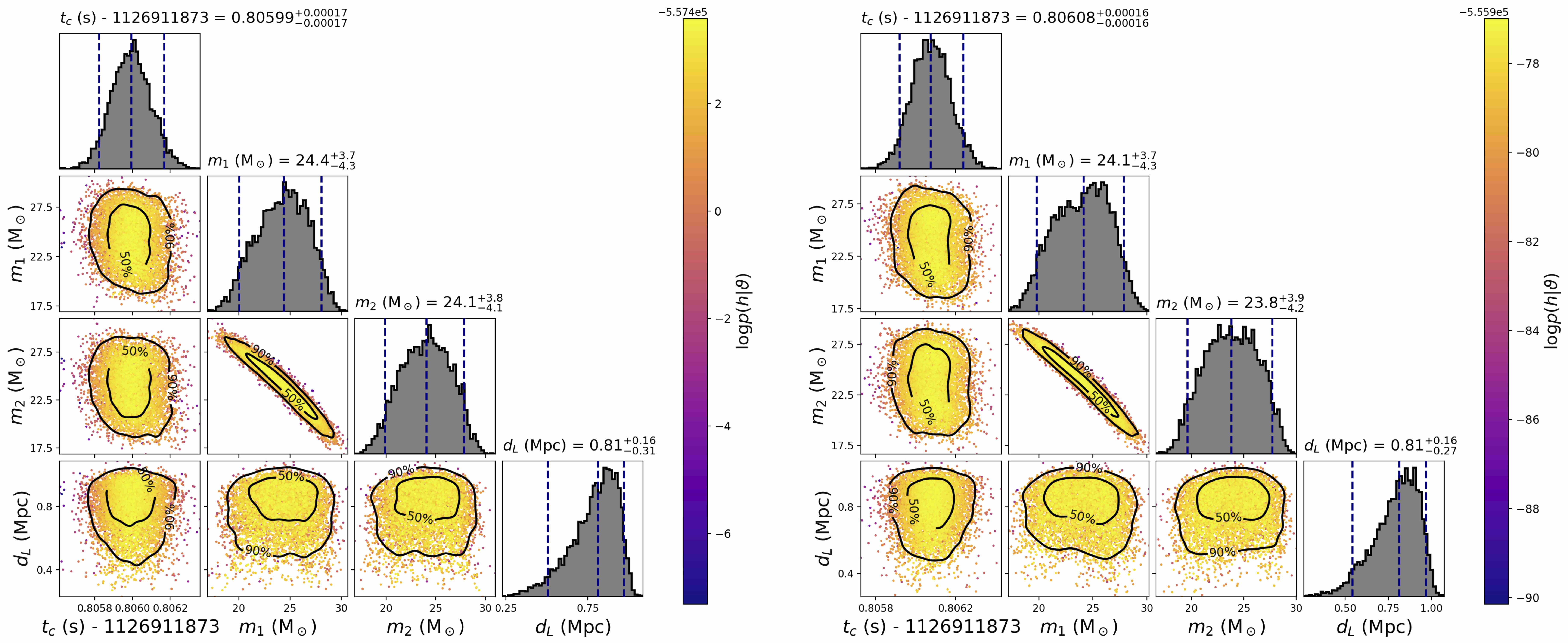}
  \caption{Posterior probability densities for trigger time, component masses, and luminosity distance before and after bilinear subtraction. The signal injection is carried out using an optimally oriented nonspinning black hole binary, of 24 solar mass each and is modeled using the frequency-domain IMRPhenomD waveform model. Matched filtering carried out using the optimal template shows an increase in SNR from 32 to 35.4\,.}
  \label{fig:inference}
\end{figure*}

\section{Results} \label{sec:results}

Results reported in this work make use of half a day of data recorded during the first observation run period of Advanced LIGO (O1). We specifically chose this period as it had the highest contamination from the above-described bilinear coupling. In Fig. \ref{fig:bilinearL_subtraction}, we show the estimated bilinear noise contribution from SR longitudinal to the strain signal for a typical data segment of an 8-minute duration and provide the cleaned spectra after carrying out time-domain subtraction. Figure \ref{fig:bilinearL_subtraction2} provides the average levels of subtraction achieved across the midfrequencies for the entire duration of the data. We see a broadband reduction of the order of 0.5 dB, with maximum suppression achieved for the sidebands seen around 320 Hz ($\sim$15 dB) and 375 Hz ($\sim$5 dB). Interestingly, the calibration line at 375 Hz is injected via the power recycling mirror, and so the fact that we are able to subtract off its sidebands with this scheme point towards a cross-coupling between PR and SR cavities. 

To assess the impact, we look at the horizon distance \citep{PhysRevD.47.2198,chen2017distance} for an optimally located and oriented binary neutron star of 1.4 solar mass each, and obtain a modest improvement close to 50 kilo-parsecs over the baseline sensitivity, as shown in subplot 1 of Fig. \ref{fig:DQ_improvements}. We also observe a 30\% reduction in the number of glitches in the data postsubtraction. As a preprocessing step towards this analysis, strain data were first whitened to enhance the high-frequency content and then subjected to a multiresolution analysis for every 1-second-long data segment. Time-frequency scalograms were then constructed via continuous wavelet transformation using analytic Morlet wavelet \citep{mallat1999wavelet,lilly2012generalized} and the pixels with excess energy within them were identified using the hierarchical algorithm for clusters and ridges algorithm \citep{Heng_2004}. Subplot 2 of Fig. \ref{fig:DQ_improvements} shows that the cumulative histogram of transients with SNR is less than a given threshold, and the effectiveness of bilinear subtraction is visibly evident. 

Channels used in the construction of the bilinear signal are not themselves sensitive to the differential arm length motion induced by GWs; hence the cleaning process described above should not, in principle, lead to the subtraction of any real GW signal. One way to check this is to ensure that the height of calibration lines remains consistent in the process. We look at a finely resolved calibration line at 434 Hz and find the relative difference to be less than 1\%. The calibrated strain data are produced by subtracting known actuation signals from the error point signal, so the calibration lines are already suppressed by a factor corresponding to the SNRs, which can vary between 10 to 50. The difference measured from bilinear filtering is hence insignificant compared to the actual height of the line. Another way to verify this is to look at an inspiral-merger-ringdown signal from a coalescing binary and observe the effect on matched filtering \citep{PhysRevD.44.3819, PhysRevD.85.122006} and parameter estimation. We carry out a software injection of an optimally oriented nonspinning black hole binary, of 24 solar mass each, using the frequency-domain IMRPhenomD \citep{PhysRevD.93.044007} phenomenological waveform model. We chose the masses accordingly to ensure that majority of the radiated energy is distributed across the midfrequency region. In this case, we observe the matched filter SNR to increase from its initial value of 32 to a value of 35.4 postfiltering. Finally, we compute the posterior probabilities for the intrinsic component masses and some of the extrinsic parameters such as trigger time and luminosity distance. The Bayesian parameter estimation is carried out using the PyCBC inference software package \citep{Biwer_2019}. We sample the posterior probability density function of the model parameters using the affine-invariant ensemble sampler, emcee \citep{goodman2010ensemble, Foreman_Mackey_2013} where the burn-in criterion for the sampler is decided based on both the autocorrelation length and maximum value of the posterior. Results shown in Fig. \ref{fig:inference} confirm that the respective posteriors for the signal parameters remain consistent with the original values.

\section{Conclusions} \label{sec:conclusions}

We presented a postoffline scheme to perform the time-domain subtraction of bilinear coupling often observed in GW detectors and demonstrated it using the GEO\,600 detector data. We made use of existing narrow-band injections to identify the coupling and showed that it is indeed possible to achieve suppression of sidebands as well as broadband noise reduction in the midfrequency range. The analysis broadened our understanding of possible contributors to the noise often seen with higher levels of circulating power. The side-band suppression demonstrated in this work could be beneficial to the signal search pipelines looking for continuous waves, but further analysis is needed to analyze the impact qualitatively. The observed glitch reduction implies such a scheme could be used in conjunction with traditional veto and gating based techniques to improve the significance of real events in transient searches. Finally, we looked at the effect on injected GW signals and observed an increase in the matched filter SNR and a consistency in the posterior probabilities of the signal parameters. Although the technique presented in this work was applied to reduce the bilinear coupling, it could be extended further to tackle higher-order couplings present in the data. Developing automated techniques to identify and subtract them off in real time would be part of future work.

\section{Acknowledgments.}

We thank the GEO collaboration for the construction of GEO\,600, and Walter Gra{\ss} for his work in keeping the interferometer in a good running state. N. M. expresses thanks to Denis Martynov and Gautam Venugopalan for their valuable suggestions on the filtering scheme. Special thanks go to Sumit Kumar and Bhooshan Gadre for their insights on Bayesian parameter estimation. The authors are grateful for support from the Science and Technology Facilities Council (STFC) Grant No. ST/L000946/1, the University of Glasgow in the United Kingdom, the Bundesministerium f\"{u}r Bildung und Forschung (BMBF), and the state of Lower Saxony in Germany. This work was partly supported by the DFG grant SFB/Transregio 7 Gravitational Wave Astronomy. This document has been assigned LIGO document number LIGO-P1900350.

\label{Bibliography}
\bibliography{references}

\end{document}